\documentstyle{article}

\begin{document}

\def\levelset{{\cal X}}
\def\fxx{{\bf x}}
\def\fxy{{\bf y}}
\def\real{I\!\!R}

\begin{center}
{\LARGE Retarded Electromagnetic Interaction and the }\\
{\LARGE Origin of Non-linear Phenomena in Optics }\\
\vskip 0.2in
{\large Mei Xiaochun}\\
\vskip 0.2in
\par
(Department of Physics, Fuzhou University, Fuzhou, 350025, China, E-mail: fzbgk@pub3.fz.fj.cn)
\end{center}
\begin{abstract}
\end{abstract}
\par
 The non-linear relation between electric polarization and electric field strength is achieved through introducing the retarded electromagnetic interactions between classical charge particles. The result agrees with the phenomenological theory in current non-linear optics, means that the non-linear phenomena in optics come from the retarded electromagnetic interaction between charged particles.
\par
The current non-linear optics describes interaction between light and materials by the half-classical method, i.e., use the classical theory of electromagnetic field to describe light's motion and quantum theory to describe interaction between photons and material particles. The theory can deal with non-linear optics well. The Maxwell equation of electromagnetic interaction in medium is
\begin{equation}
\nabla\times\vec{E}={{\partial\vec{B}}\over{\partial{t}}}~~~~~~~~~~~~~~~~~~\nabla\times\vec{H}=\vec{J}_f+{{\partial\vec{D}}\over{\partial{t}}}
\end{equation}
\begin{equation}
\nabla\cdot\vec{D}=\rho_f~~~~~~~~~~~~~~~~~~~~~~\nabla\cdot\vec{B}=0
\end{equation}
The experiments show that for general non-iron electric medium and non-iron magnetic medium, under the condition of weak field, the relations between electric polarization $\vec{P}$ and electric field strength, as well as magnetic polarization $\vec{M}$ and magnetic field strength are linear, connected by the so-called constructive equations
\begin{equation}
\vec{D}=\varepsilon\vec{E}=\varepsilon_0\vec{E}+\vec{P}~~~~~~~~~~~~~~~~~~\vec{B}=\mu\vec{H}=\mu_0(\vec{H}+\vec{M})
\end{equation}
\begin{equation}
\vec{P}=\varepsilon_0\chi_e\vec{E}~~~~~~~~~~~~~~~~~~~~~~~\vec{M}=\chi_m\vec{H}
\end{equation}
But the experiments also show that in the strong fields, the relations between $\vec{P}$ and $\vec{E}$, as well as $\vec{M}$ and $\vec{H}$ are non-linear and vary complex. This kind of relations can't be deduced from theory at present. We now get them by phenomenological hypothesis. For example, in the non-linear optics, the relation between electric polarization $\vec{P}$ and electric field strength is supposed to be $^{(1)}$
\begin{equation}
\vec{P}=\varepsilon_0(\chi^1_e\vec{E}+\vec{\chi}^2_e\cdot\vec{E}\vec{E}+\vec{\vec{\chi}}^3_e:\vec{E}\vec{E}\vec{E}\cdot\cdot\cdot)
\end{equation}
Here $\chi^i_e$ is polarization tensors. By means of the formula and the Maxwell equation, so much non-linear phenomena in optics can be described well. It is proved blow that after the retarded electromagnetic interaction between charged particles are considered, we can obtain Eq.(5) easy and shows that the non-linear phenomena in optics come from the retarded electromagnetic interaction between charged particles.
\par
Let $t',\vec{r}',\vec{v}'$ and $\vec{a}'$ represent retarded time, coordinate, velocity and acceleration, $t,\vec{r},\vec{v}$ and $\vec{a}$ represent non-retarded time, coordinate, velocity and acceleration. A particle with charge $q_i$, velocity $\vec{v}'_j$ and acceleration $\vec{a}'_j$ at space point $\vec{r}'_j(t')$ and time $t'$ would cause retarded potentials as follows at space point $\vec{r}_i(t)$ and time $t$ 
\begin{equation}
\varphi'_{ij}={{q_j}\over{4\pi\varepsilon_0(1-{{\vec{v}'_j\cdot\vec{n}_{ij}}\over{c}})r'_{ij}}}~~~~~~~~~~~~~~\vec{A}'_{ij}={{q_j\vec{v}'_j}\over{4\pi\varepsilon_0(1-{{\vec{\nu}'_j\cdot\vec{n}_{ij}}\over{c}})r'_{ij}}}
\end{equation}
In the formula $\vec{r}'_{ij}(t,t')=\vec{r}'_i(t)-\vec{r}'_j(t')$, $r'_{ij}=\mid\vec{r}'_{ij}\mid$, $\vec{n}_{ij}=\vec{r}'_{ij}/{r}'_{ij}$. When the particle's speed $v<<{c}$, the retarded distance $r'_{ij}(t')$ at retarded time $t'$ can be replaced approximately by non-retarded distance $r_{ij}(t)$, i.e., we can let
\begin{equation}
t'=t-r'_{ij}(t')/{c}\rightarrow{t}-r_{ij}(t)/{c}
\end{equation}
\begin{equation}
r'_{ij}(t,t')=\mid\vec{r}_i(t)-\vec{r}'_j(t')\mid=\mid\vec{r}_i(t)-\vec{r}'_j[t-r'_{ij}(t')/{c}]\mid\rightarrow\mid\vec{r}_i(t)-\vec{r}'_j[t-r_{ij}(t)/{c}]\mid=r'_{ij}(t)
\end{equation}
It is noted that $r'_{ij}(t)\neq{r}_{ij}(t)$, for $r'_{ij}(t)$ is the approximate retarded distance, but $r_{ij}(t)$ is not the retarded distance. In this case, we can develop retarded quantities into series in light of small quantity $r_{ij}/{c}$. By relation $\vec{v}_j=d\vec{r}_j/{d}t$, we get $^{(2)}$ 
$$\vec{r}'_{ij}(t,t)\simeq\vec{r}_i(t)-\vec{r}'_j(t-r_{ij}/{c})=\vec{r}_i(t)-\vec{r}_j(t)+{{\vec{v}_j(t)}\over{c}}{r}_{ij}(t)-{{\vec{a}_j(t)}\over{2c^2}}r^2_{ij}(t)+{{\dot{\vec{a}}_j(t)}\over{6c^3}}{r}^3_{ij}(t)+\cdot\cdot\cdot$$
\begin{equation}
=\vec{r}_{ij}(t)+{{\vec{v}_j(t)}\over{c}}{r}_{ij}(t)-{{\vec{a}_j(t)}\over{2c^3}}r^2_{ij}(t)+{{\dot{\vec{a}}_j(t)}\over{6c^3}}r^3_{ij}(t)+\cdot\cdot\cdot
\end{equation}
\begin{equation}
\vec{v}_j(t')\simeq\vec{v}_j(t)-{{\vec{a}_j(t)}\over{c}}{r}_{ij}(t)+{{\dot{\vec{a}}_j(t)}\over{2c^3}}{r}^2_{ij}(t)+\cdot\cdot\cdot~~~~~~~~~~~\vec{a}'_j(t')\simeq\vec{a}_j(t)-{{\dot{\vec{a}}_j(t)}\over{c}}{r}_{ij}(t)+\cdot\cdot\cdot
\end{equation}
\par
Mediums are composed of atoms and molecules, and atoms and molecules can be regarded as electrical dipoles composed of two particles with charges $\pm{q}$. Suppose the distance between two charges is $r_{ij}(t)$ at time $t$, the electrical dipole moment is $\vec{P}_j$. The direction of $\vec{P}_j=q_j\vec{r}_{ij}$ is from the point the $j$ -particle located pointing to the mass center point $\vec{r}_i$ of electric dipole. By considering the retarded effect of interaction, according to Eq.(9), the retarded electrical dipole moment becomes approximately
\begin{equation}
\vec{P}'_j=q_j\vec{r}'_{ij}=q_j(\vec{r}_{ij}+{{\vec{\nu}_j}\over{c}}{r}_{ij}-{{\vec{a}_j}\over{2c^2}}r^2_{ij}+{{\dot{\vec{a}}_j}\over{6c^3}}r^3_{ij})
\end{equation}
$\vec{v}_j(t)$£¬$\vec{a}_j(t)$ and $\dot{\vec{a}}_j(t)$ are the -particle's velocity, acceleration and acceleration of acceleration individually. If there is only one electrical dipole in space, the directions of $\vec{r}_{ij}$, $\vec{v}_j$, $\vec{a}_j$ and $\dot{\vec{a}}_j$ are on the same straight line. In this case, the subscript can be removed and Eq.(11) can be written as
\begin{equation}
P'=q(r+{v\over{c}}r-{{a}\over{2c^2}}r^2+{{\dot{a}}\over{6c^3}}r^3+\cdot\cdot\cdot)
\end{equation}
When there exists the action of external electric field $E=E_0\sin(\omega{t}+\theta'_0)$, the motion equation of electrical dipole is
\begin{equation}
{{d^2r}\over{dt^2}}+2\beta{{dr}\over{dt}}+\omega^2_0{r}={{qE_0}\over{m}}\sin(\omega{t}-\theta'_0)
\end{equation}
The second item on the right side is damping force (radiation damping force and damping force caused by collisions between particles). When $\omega_2>\beta$, the solution of Eq.(13) is $^{(3)}$ 
\begin{equation}
r=A\sin(\omega{t}-\theta_0)+Be^{-\beta{t}}\sin(\sqrt{\omega^2_0-\beta^2{t}}-\delta)
\end{equation}
\begin{equation}
A={{qE_0}\over{m\sqrt{(\omega^2_0-\omega^2)^2+(2\beta\omega)^2}}}~~~~~~~~~~~~\theta_0=\theta'_0+tg^{-1}({{\beta\omega}\over{\omega^2_0-\omega^2}})
\end{equation}
When $t$ is big enough, we have $e^{-\beta{t}/{2}}\rightarrow{0}$ and the second item on the right side of Eq.(14) can be omitted. In this case, we have
\begin{equation}
v=A\omega\cos(\omega{t}-\theta_0)~~~~~~~~~~a=-A\omega^2\sin(\omega{t}-\theta_0)~~~~~~~~~~~\dot{a}=-A\omega^3\cos(\omega{t}-\theta_0)
\end{equation}
Put them into Eq.(12), we get
$$P'=qA\{\sin(\omega{t}-\theta_0)+{{\omega}\over{c}}\sin(\omega{t}-\theta_0)\cos(\omega{t}-\theta_0)+{{\omega^2}\over{2c^2}}{\sin}^3(\omega{t}-\theta_0)$$
$$-{{{\omega}^3}\over{6c^3}}{\sin}^3(\omega{t}-\theta_0)\cos(\omega{t}-\theta_0)+\cdot\cdot\cdot\}$$
$$=qA\{(1+{{3\omega^2}\over{8c^2}})\sin(\omega{t}-\theta_0)+{{\omega}\over{2c}}(1-{{\omega^2}\over{24c^2}})\sin(2\omega{t}-2\theta_0)-{{\omega^2}\over{8c^2}}\sin(3\omega{t}-3\theta_0)$$
\begin{equation}
+{{{\omega}^3}\over{48c^3}}\sin(4\omega{t}-4\theta_0)+\cdot\cdot\cdot\}
\end{equation}
It is obvious that after the retarded interaction is considered, the waves of $2,3,4\cdot\cdot\cdot$ multiple frequencies would appear in the vibrations of electrical dipoles. If two electric fields with different frequencies are imported with
\begin{equation}
E=E_1\sin(\omega_1{t}-\theta'_1)+E_2\sin(\omega_2{t}-\theta'_2)
\end{equation}
similar results can be obtained
\begin{equation}
r=A_1\sin(\omega_1{t}-\theta_1)+A_2\sin(\omega_2{t}-\theta_2)
\end{equation}
\begin{equation}
\nu=A_1\omega_1\cos(\omega_1{t}-\theta_1)+A_2\omega_2\cos(\omega_2{t}-\theta_2)
\end{equation}
\begin{equation}
a=-A_1\omega^2_1\sin(\omega_1{t}-\theta_1)-A_2\omega^2_2\sin(\omega_2{t}-\theta_2)
\end{equation}
\begin{equation}
\dot{a}=-A_1\omega^3_1\cos(\omega_1{t}-\theta_1)-A_2\omega^2_2\cos(\omega_2{t}-\theta_2)
\end{equation}
In the formula, $A_i=qE_0/{m}\sqrt{(\omega^2_0-\omega^2_i)^2+(2\beta\omega_i)^2}$, $\theta_i=\theta'_i+tg^{-1}[2\beta\omega/(\omega^2_0-\omega^2)]$. Put them into Eq.(12), we get
$$P'=q\{[A_1+{{A^2_1\omega_1}\over{2c}}+A_1{A}^2_2({{\omega^2_1}\over{2}}+\omega^2_2)]\sin(\omega_1{t}-\theta_1)+[A_2+{{A^2_2\omega_2}\over{2c}}+A^2_1{A}_2(\omega^2_1+{{\omega^2_1}\over{2}})]\sin(\omega_2{t}-\theta_2)$$
$$+{{A_1{A}_2}\over{2c}}[(\omega_1+\omega_2)\sin((\omega_1+\omega_2)t-(\theta_1+\theta_2))+(\omega_1-\omega_2)\sin((\omega_2-\omega_1)t-(\theta_2-\theta_1))]$$
$$+{{A^3_1\omega^2_1}\over{4c}}[\sin(3\omega_1{t}-3\theta_1)-3\sin(\omega_1{t}-\theta_1)]+{{A^3_2\omega^2_2}\over{4c}}[\sin(3\omega_2{t}-3\theta_2)-3\sin(\omega_2{t}-\theta_2)]$$
$$-A_1{A}^2_2({{{\omega}^2_1}\over{4}}+\omega^2_2)[\sin((\omega_1+2\omega_2)t-(\theta_1+2\theta_2))+\sin((\omega_1-2\omega_2)t-(\theta_1-2\theta_2))]$$
\begin{equation}
-A^2_1{A}_2(\omega^2_1+{{{\omega}^2_2}\over{4}})[\sin((\omega_2+2\omega_1)t-(\theta_2+2\theta_1))+\sin((\omega_2-2\omega_1)t-(\theta_2-2\theta_1))]+\cdot\cdot\cdot\}
\end{equation}
Besides multiple frequencies, sum frequencies and different frequencies would appear.
\par
Because $v=A\omega\cos(\omega{t}-\theta_0)=A\omega\sin(\omega{t}-\theta_0-\pi\prime{2})$, there is a difference of phase $\pi/{2}$ between $v/\omega$ and $r$. There is also a difference of phase $\pi/{2}$ between $\dot{a}/\omega^3$ and $r$. If these differences are omitted, we can let $v\simeq\omega{r}$, $\dot{a}\simeq\omega^3{r}$. Considering $a=-\omega^2{r}$, put the relations into Eq.(12), we get
\begin{equation}
P'\simeq{q}(r+{{\omega}\over{c}}{r}^2+{{\omega^2}\over{2c^2}}{r}^3+{{\omega^3}\over{6c^3}}{r}^4+\cdot\cdot\cdot)
\end{equation}
According to Eq.(14), when $t$ is large enough and the differencd of phase is omitted, the displacement of dipole directs radio to electric field's strength, i.e., $r=\alpha{E}$, $\alpha=q/{m}\sqrt{(\omega^2_0-\omega^2)^2+(2\beta\omega)^2}$. For the vibrations of atoms in crystal lattice, we have $\omega>>\omega_0$ and $\omega>>\beta$, so $\alpha\simeq{q}/{m}\omega^2$. Put the relation into the formula above, we get
\begin{equation}
P'\simeq{q}\alpha(E+{{\omega\alpha}\over{c}}{E}^2+{{\omega^2\alpha^2}\over{2c^2}}{E}^3+{{\omega^3\alpha^3}\over{6c^3}}{E}^4+\cdot\cdot\cdot)
\end{equation}
This is just the non-linear relation between electrical polarization and electric field strength. Suppose the wavelength of incident wave is $\lambda=4\times{10}^{-7}{M}$, the dipole vibrating in lattice is hydrogen with mass $m$, we get
\begin{equation}
{{\omega\alpha}\over{c}}={{q}\over{cm\omega}}\simeq{3.14}\times{10}^{-10}{M}/{V}
\end{equation}
The radio of parameters between the first item and second item is about $10^{-10}$. This value coincides with experiments $^{(4)}$. Eq.(27) can be written as
\begin{equation}
P'\simeq\varepsilon_0(\chi^1_e{E}+\chi^2_e{{\omega\alpha}\over{c}}{E}^2+{{\omega^2\alpha^2}\over{2c^2}}{E}^3+{{\omega^3\alpha^3}\over{6c^3}}{E}^4+\cdot\cdot\cdot)
\end{equation}
\begin{equation}
\chi^1_e={{q\alpha}\over{\varepsilon_0}}~~~~~~~~~\chi^2_e={{q\omega\alpha^2}\over{\varepsilon_0{c}}}~~~~~~~~~\chi^3_e={{q\omega^2\alpha^3}\over{2\varepsilon_0{c}^2}}~~~~~~~~~\chi^4_e={{q\omega^3\alpha^4}\over{6\varepsilon_0{c}^3}}
\end{equation}
But it should be noted that the results exists phase differences between on Eq.(29) and (12) and Eq.(12) is fundamental.
\par
In the practical systems, there exist many dipoles and interactions between dipoles. Besides vibrations, dipoles may rotate and shift, so the directions of $\vec{r}_{ij}$, $\vec{v}_j$, $\vec{a}_j$ and $\dot{\vec{a}}_j$ are not on the same straight lines in general. By taking sum and let $q_j\alpha_j=\varepsilon_0\chi^1_{ej\sigma}$, $q_j\omega_j\alpha^2_j/{c}=\varepsilon_0\chi^2_{ej\sigma}\cdot\cdot\cdot$, the total electrical dipole moment of medium can be written as the form of partial quantities 
\begin{equation}
P_{\sigma}=\sum^N_{j=1}{P}'_{j\sigma}=\varepsilon\sum^N_{j=1}(\chi^1_{ej\sigma}{E}+\chi^2_{ej\sigma}{E}^2+\chi^3_{ej\sigma}{E}^3+\cdot\cdot\cdot)
\end{equation}
Because the model of dipole is only a kind of approximation, by considering interactions between particles as well as by considering other complex factors, in general situation, the non-linear relation between electrical polarization and electric field strength should be written as the form of product form of tensors
\begin{equation}
\vec{P}=\varepsilon_0(\chi^1_e\vec{E}+\vec{\chi}^2_e\cdot\vec{E}\vec{E}+\vec{\vec{\chi}}^3_e:\vec{E}\vec{E}\vec{E}+\cdot\cdot\cdot)
\end{equation}
This is just the foundational relation used in non-linear optics.
\par
In order to explain the origin of non-linear phenomena in optics or Eq.(32), the non-linear oscillator model is discussed at present. The motion equation of oscillator is written as $^{(1)}$ 
\begin{equation}
{{d^2r}\over{dt^2}}+2\beta{{dr}\over{dt}}+\omega^2_0{r}+Dr^2={F\over{m}}
\end{equation}
In the formula, $mDr^2$ is the non-linear force to cause non-linear effects. However, it should be seen that the origin of this item is still unclear. Where it comes from? Why is takes such form, not $mDr^{1/{2}}$ or $mDr^3$ and so on. The current theory can't provide clear explanation. Meanwhile, the value of parameter $D$ can't be decided by theory. Quantum mechanics provides a method of perturbation approximation to calculate polarizability, but does not yet provide explanation for the origin of non-linear phenomena. So it is proper to consider that the origin of non-linear in optics comes from the retarded electromagnetic interaction.
\par
Because $\vec{v}\rightarrow-\vec{v}$, $\vec{a}\rightarrow\vec{a}$, $\dot{\vec{a}}\rightarrow\dot{\vec{a}}$ under time reversal, so $\vec{r}'$ shown in Eq.(11) can't keep unchanged under time reversal. In this way, it can be said that most of non-linear phenomenon in optics violate symmetry of time reversal.
\\
\\
Reference
\\
(1) Zuo Chongpei, Non-linear Physics, Tianjing Science and Technology Publishing House, 267£¬271
£¨1995£©.
\\
(2) Cao Changqi, Electrodynamic, People Education House, 240£¨1979£©.
\\
(3) Zhuo Yanbo, Theoretical Mechanics, Jiangshu Since and Technology Publishing House, 107£¨1979£©.
\\
(4) J£®A£®Wan-Vechten£¬Phys. Rev.,183, 709 (1969).
\end{document}